\documentstyle[aps,prl,twocolumn,epsfig,floats]{revtex}
\title{ Evidence of  $\kappa$ particle in
$J/\psi \to \bar{K}^*(892)^0 K^+ \pi^-$
\thanks{Data analysized were taken prior to the participation
of U.S. members and Japanese members of BES Collaboration.}}
 
\author{\small   (BES Collaboration) \\
\begin{minipage}{16cm}
\small
J.~Z.~Bai$^1$,        Y.~Ban$^{8}$,         J.~G.~Bian$^1$,
X.~Cai$^{1}$,          J.~F.~Chang$^1$,
H.~F.~Chen$^{13}$,    H.~S.~Chen$^1$,
Jie~Chen$^{7}$,       J.~C.~Chen$^1$,     
Y.~B.~Chen$^1$
% ...
\\
\small
\begin{center}
$^1$ Institute of High Energy Physics, Beijing 100039, People's Republic of
     China\\
$^2$ China Center of Advanced Science and Technology, Beijing 100080,
     People's Republic of China\\
$^3$ Henan Normal University, Xinxiang 453002, People's Republic of China\\
$^4$ Huazhong Normal University, Wuhan 430079, People's Republic of China\\
$^5$ Hunan University, Changsha 410082, People's Republic of China\\
$^{6}$ Liaoning University, Shenyang 110036, People's Republic of China \\
$^{7}$ Nankai University, Tianjin 300071, People's Republic of China\\
$^{8}$ Peking University, Beijing 100871, People's Republic of China\\
$^{9}$ Shandong University, Jinan 250100, People's Republic of China\\
$^{10}$ Shanghai Jiaotong University, Shanghai 200030, 
        People's Republic of China\\
$^{11}$ Sichuan University, Chengdu 610064,
        People's Republic of China\\                                    
$^{12}$ Tsinghua University, Beijing 100084, 
        People's Republic of China\\
$^{13}$ University of Science and Technology of China, Hefei 230026,
        People's Republic of China\\
$^{14}$ Wuhan University, Wuhan 430072, People's Republic of China\\
$^{15}$ Zhejiang University, Hangzhou 310028, People's Republic of China\\
\vspace{0.4cm}
$^{\ast}$ Visiting professor to University of Michigan, Ann Arbor, 
MI 48109 USA
\end{center}
\end{minipage}
}

\begin{document}
%\date{}
\maketitle
\vskip -0.5cm

%\normalsize

\begin{abstract}
Based on a $J/\psi$ data sample of $7.8 \times 10^6$ 
events at BESI,   
the decay of $ J/\psi \to \bar{K}^*(892)^0 K^+ \pi^-$
is studied and a low mass enhancement, which is believed not 
coming from the phase space effect or background, is visible 
in the $K^+ \pi^-$ invariant mass spectrum recoiling against
$\bar{K}^*(892)^0$.
Partial wave analysis of this channel favors this low 
mass enhancement being a broad $0^{+}$ resonance with the mass 
and width of $771^{+164}_{-221}\pm 55$ MeV/c$^2$ and
$220^{+225}_{-169} \pm 97$ MeV/c$^2$, respectively. 

\vspace{3\parskip}

\end{abstract}
%-------------------------------------------------------
%\clearpage

The early analyses of $\pi \pi$/$\pi K$ scattering data 
showed no pole at the low mass region\cite{1,2,3}. 
However, chiral theory predicted the existence of 
$\sigma$ and $\kappa$\cite{7,5,6}. 
Re-analyses of the $\pi \pi$ and $\pi K$ scattering data
then showed an evidence for the existence of the $\sigma$ and
$\kappa$ particle. 
The pole position of $\kappa$ was determined 
to be \cite{7} 
$\sqrt{s_{\kappa}}=(0.875 \pm 0.075) - i(0.335 \pm 0.110)GeV$.
Motivated by the $1/N_c$ expansion, 
a simple model was proposed, which included
a light strange scalar meson $\kappa$ with the pole position at
$\sqrt{s_{\kappa}}=(0.911 - 0.158i)GeV$\cite{5}. 
A method of resummation of the
$\chi PT$ series based on the expansion of $T^{-1}$
was studied\cite{6}, which gave the 
mass and width of $\kappa$ particle as
$m_{\kappa} = 770$ MeV/c$^2$, $\Gamma_{\kappa} = 500$ MeV/c$^2$.
All evidences mentioned above were from $\pi \pi$ 
and $\pi K$ scattering.
Recently, an evidence for $\sigma$ and
$\kappa$ has been reported in the production 
process by  experiments\cite{66,12,e791}.
In this work, we analyze
the data of the $K \pi$ system in 
the production process $ J/\psi \to \bar{K}^*(892)^0 K^+ \pi^-$
to search for the $\kappa$. \\ 

A total of 
$7.8 \times 10^6$  $J/\psi$ events, which were accumulated 
at the Beijing Spectrometer\cite{11}, is used to analyze
$ J/\psi \to \bar{K}^*(892)^0 K^+ \pi^-$ channel.
The preselection requires that the candidate events 
have four good charged tracks, which are in the 
polar angle region of $\mid{\cos{\theta}}\mid<0.84$, 
with a total of zero
net charge, without any isolated photon.  
The tracks must have good helix fit 
in the Main Drift Chamber(MDC) and have the vertices
within an interaction region of $R_{xy}<0.020$m and
${\mid{z}\mid}<0.20$m. Isolated photons, which are
not associated with charged tracks,  are identified 
by requiring that the energy deposited in the 
Barrel Shower Counter (BSC) is greater 
than 50 MeV, the  shower starts before the layer 6
and the angle between the direction at the first layer
of BSC and the developing direction of the cluster is less 
than $30^\circ$.  
All the surviving events are submitted to the
4-constraints kinematic fits for the following 5 cases:
$K^+ K^- \pi^+ \pi^-$, $\pi^+ \pi^- \pi^+ \pi^-$,
$K^+ K^- K^+ K^-$, $K^+ \pi^- \pi^+ \pi^-$ and
$\pi^+ K^- \pi^+ \pi^-$.  
For each case,  all the possible combinations of 
4 charged tracks are tried and the one which
has the least $\chi^2$ value is chosen.
To select $K^+ K^- \pi^+ \pi^-$ events, 
it is required 
that $\chi^2_{J/\psi \to  K^+ K^- \pi^+ \pi^-} < 30$
and $\chi^2_{J/\psi \to  K^+ K^- \pi^+ \pi^-} <
\chi^2_{J/\psi \to \pi^+ \pi^- \pi^+ \pi^-}$,
$\chi^2_{J/\psi \to K^+ K^- K^+ K^-}$, 
$\chi^2_{J/\psi \to K^+ \pi^- \pi^+ \pi^-}$
and $\chi^2_{J/\psi \to \pi^+ K^-   \pi^+ \pi^-}$. 
The information of
Time-of-Flight counter(TOF) is used for the
particle identification. All particles with momentum
below 500 MeV/c should have consistent particle 
identification with $K^+ K^- \pi^+ \pi^-$ assumption.
In order to remove background events
from $J/\psi \to \phi \pi^+ \pi^-$ and $J/\psi \to 
\bar{K}^*(892)^0 K_s^0$, 
events with invariant mass of $K^+ K^-$  satisfying
$|M_{K^+ K^-}-1.02|< 0.02$ GeV/c$^2$ or $R_{xy}$ 
of any track  greater than 0.010m are rejected. 
Finally, it is  required that  
$ | M_{K^- \pi^+} - 0.892 | \le 0.06$ GeV/c$^2$ to 
select $J/\psi \to \bar{K}^*(892)^0 K^+ \pi^-$.\\

\begin{figure}[htbp]
\begin{center}
{\mbox{\epsfig{file=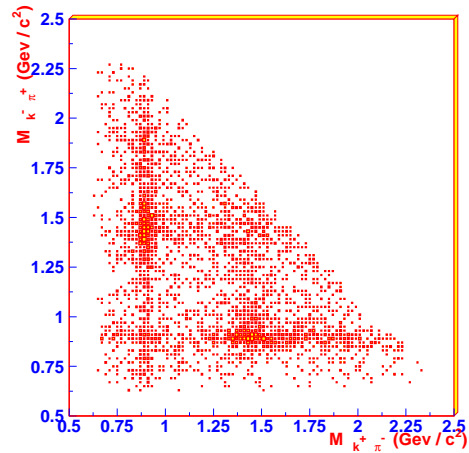,height=2.6in,width=2.6in}}}
\end{center}
\caption[]{ Scatter plot of $M_{K^+ \pi^-}$ vs. $M_{K^- \pi^+}$. A
horizontal band and a vertical band can be clearly seen, which
correspond to  $J/\psi \to \bar{K}^*(892)^0 K^+ \pi^-$ and
$J/\psi \to {K}^*(892)^0 K^- \pi^+$ respectively.
}
\label{w002}
\end{figure}

\begin{figure}[htbp]
\begin{flushleft}
{\mbox{\epsfig{file=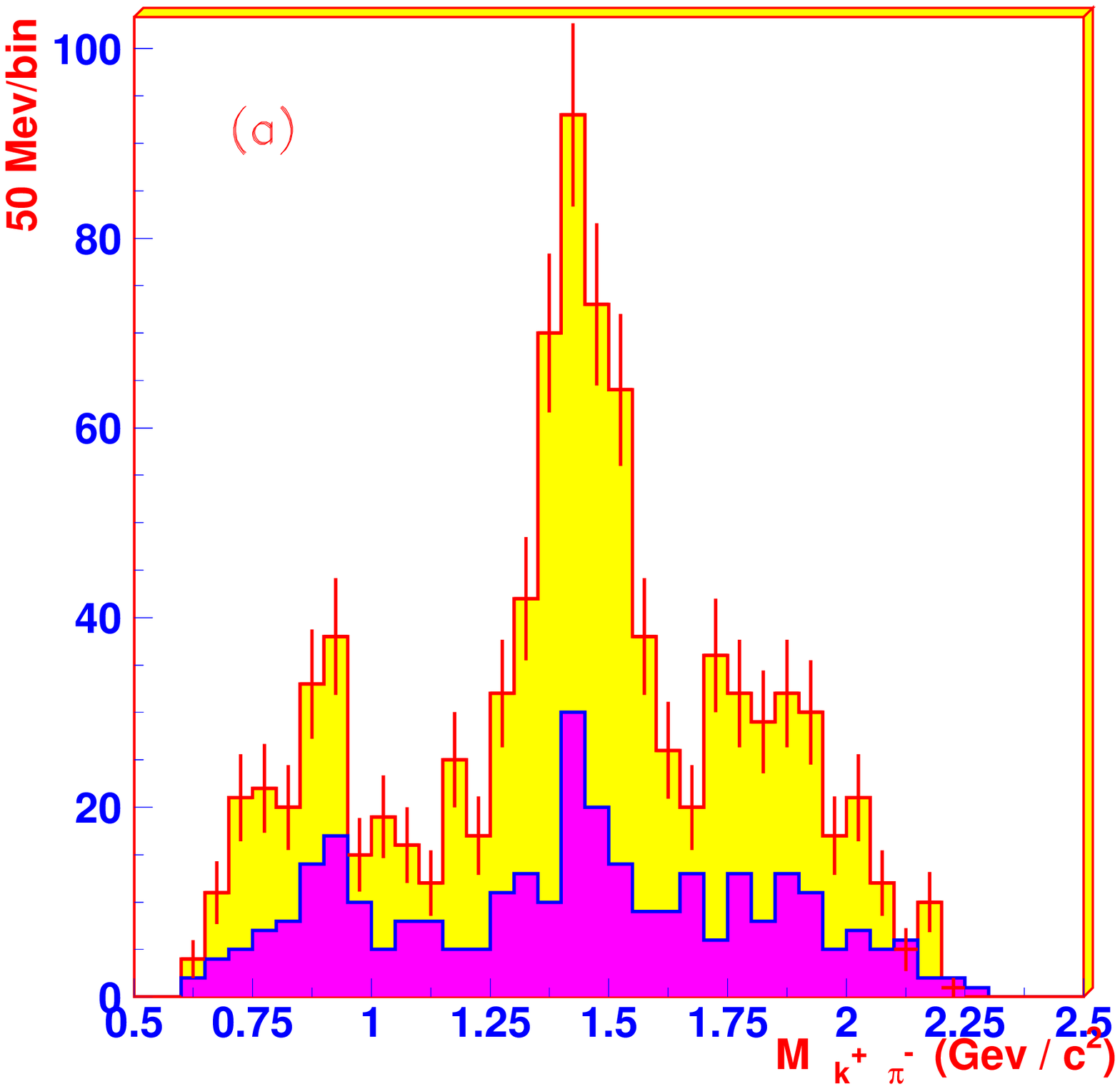,height=1.7in,width=1.7in}}}
\end{flushleft}
\vspace {-1.9in}
\begin{flushright}
{\mbox{\epsfig{file=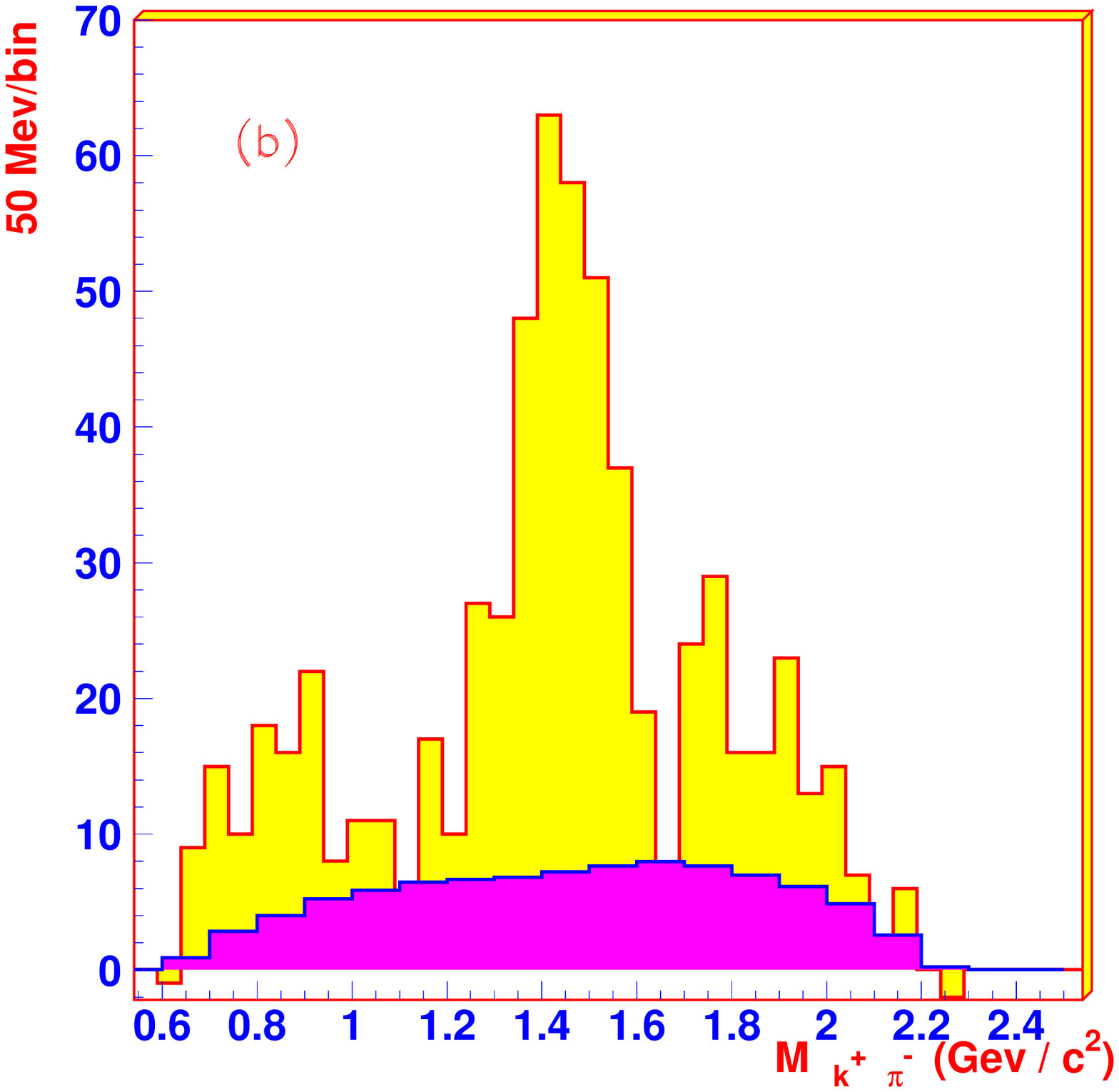,height=1.7in,width=1.7in}}}
\end{flushright}
\caption[]{ (a) The invariant mass spectrum of 
	$K^+ \pi^-$ recoiling against $\bar{K}^*(892)^0$. 
        The histogram with error bars is the data.
	The heavy shaded region shows the $\bar{K}^*(892)^0$ 
	side-band structure.  
	(b) The invariant 
	mass spectrum of $K^+ \pi^-$
        after $\bar{K}^*(892)^0$ side-band subtraction.
        The heavy shaded region shows the phase space shape.}  
\label{w031} 
\end{figure}

\begin{figure}[htbp]
\centerline{\epsfig{file=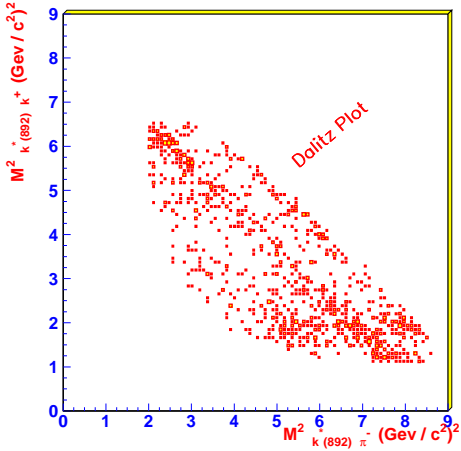,height=2.6in,width=2.6in}}
\caption[]{ Daliz plot. The top slope band corresponds
to kappa and the second slope band 
corresponds to the peak at 1430 MeV.
The horizontal band corresponds to $J/\psi \to K_1(1410) K$
and $J/\psi \to K_1(1270) K$ . }
\label{w04} 
\end{figure}

Fig.\ref{w002} shows the scatter    plot of $M_{K^+ \pi^-}$
vs. $M_{K^- \pi^+}$ after the above cuts.
Two bands of
$\bar{K}^*(892)^0$ and ${K}^*(892)^0$, corresponding
to $J/\psi \to \bar{K}^*(892)^0 K^+ \pi^-$
and $J/\psi \to {K}^*(892)^0 K^- \pi^+$ respectively,
can be clearly seen.
The invariant mass spectrum of $K \pi$
system is shown in Fig.\ref{w031}(a).
Three structures  can be clearly seen,  with
one peak at about 1.43 GeV/c$^2$,
a narrow $K^*(892)^0$ signal and a broad enhancement
lying under $K^*(892)^0$ peak below 1.1 GeV/c$^2$.
The heavy shaded region in Fig.\ref{w002} shows
the $\bar{K}^*(892)^0$ side-band structure.
A detailed Monte Carlo study is performed and
the background from other $J/\psi$
decay channels is found to be very small.
A narrow $K^*(892)^0$ peak can also be seen
in the spectrum  of $\bar{K}^*(892)^0$ side-band.
After the side-band subtraction, the narrow $K^*(892)^0$
peak disappears, which indicates that it is not from
the channel we are analyzing,  but
from the charge conjugate channel $J/\psi \to
{K}^*(892)^0 K^- \pi^+$. The invariant
mass spectrum of $K^+ \pi^-$ after the side-band subtraction
is plotted in Fig.\ref{w031}(b), where the shaded region shows
the phase space shape. The low mass broad enhancement,
which is different from the shape of the phase space,
is still clear.
Fig.\ref{w04} shows the Daliz plot, where
two slope bands and one horizontal band are visible.
The top slope band corresponds to the low mass 
enhancement which is the subject of this study,
and the second slope band corresponds to the peak at 1430 MeV.
The horizontal band corresponds to $J/\psi \to K_1(1410) K$
and $J/\psi \to K_1(1270) K$ .\\

To further study the low mass enhancement, 
a Partial Wave Analysis(PWA) is  performed for 
the $K^+ \pi^-$ invariant mass spectrum in
$J/\psi \to \bar{K}^*(892)^0 K^+ \pi^-$ channel. 
The amplitudes of the partial waves are constructed
by the covariant helicity coupling amplitude\cite{14,wun}
and the low mass enhancement is treated as a s-channel 
resonance. The analysis method and the theoretical formula
is reported in literature \cite{wun}. 
The total differential cross section which
describes the whole decay process is
$$
\frac{{\rm d} \sigma}{{\rm d} \Phi} =
\sum_{m} \sum_{\lambda} | \sum_{R_i}  \sum_{\mu}  
A_{R_i}(m, \lambda, \mu, M_i, \Gamma_i, \alpha)  |^2
+ BG ,
$$
where $A_{R_i}(m, \lambda, \mu, M_i, \Gamma_i, \alpha)$
is the helicity amplitude for the resonance $R_i$\cite{wun},
$m$ is the magnetic quantum number of $J/\psi$,
$\lambda$ and $\mu$ are  helicities of $\bar{K}^*(892)^0$
and the resonance $R_i$ respectively, $M_i$ and $\Gamma_i$
are the mass and width of resonance $R_i$,
$\alpha$ relates to the magnitude of each helicity
amplitude and $BG$ represents background contribution. 
In the fit, $M_i$, $\Gamma_i$ and $\alpha$
are free parameters to be determined.
The probability density function $f(M_i, \Gamma_i, \alpha)$ 
that describes the data sample  is
$$
f(M_i, \Gamma_i, \alpha) = \frac{d\sigma}{d\Phi} / \sigma
$$
where $\sigma$ is the total cross section of this process 
which is given by Monte Carlo integration.
The likelihood function ${ L}$ is given by the joint 
probability density of all experimental data
$$
{L}={\displaystyle \prod_{i=1}^{N_{event}}
f(M_i, \Gamma_i, \alpha)},
$$
and the log likelihood $S = - {\rm ln} { L}$ is minimized
in our analysis by varying parameters 
$M_i$, $\Gamma_i$ and $\alpha$. \\

\begin{figure}[htbp]
\begin{flushleft}
{\mbox{\epsfig{file=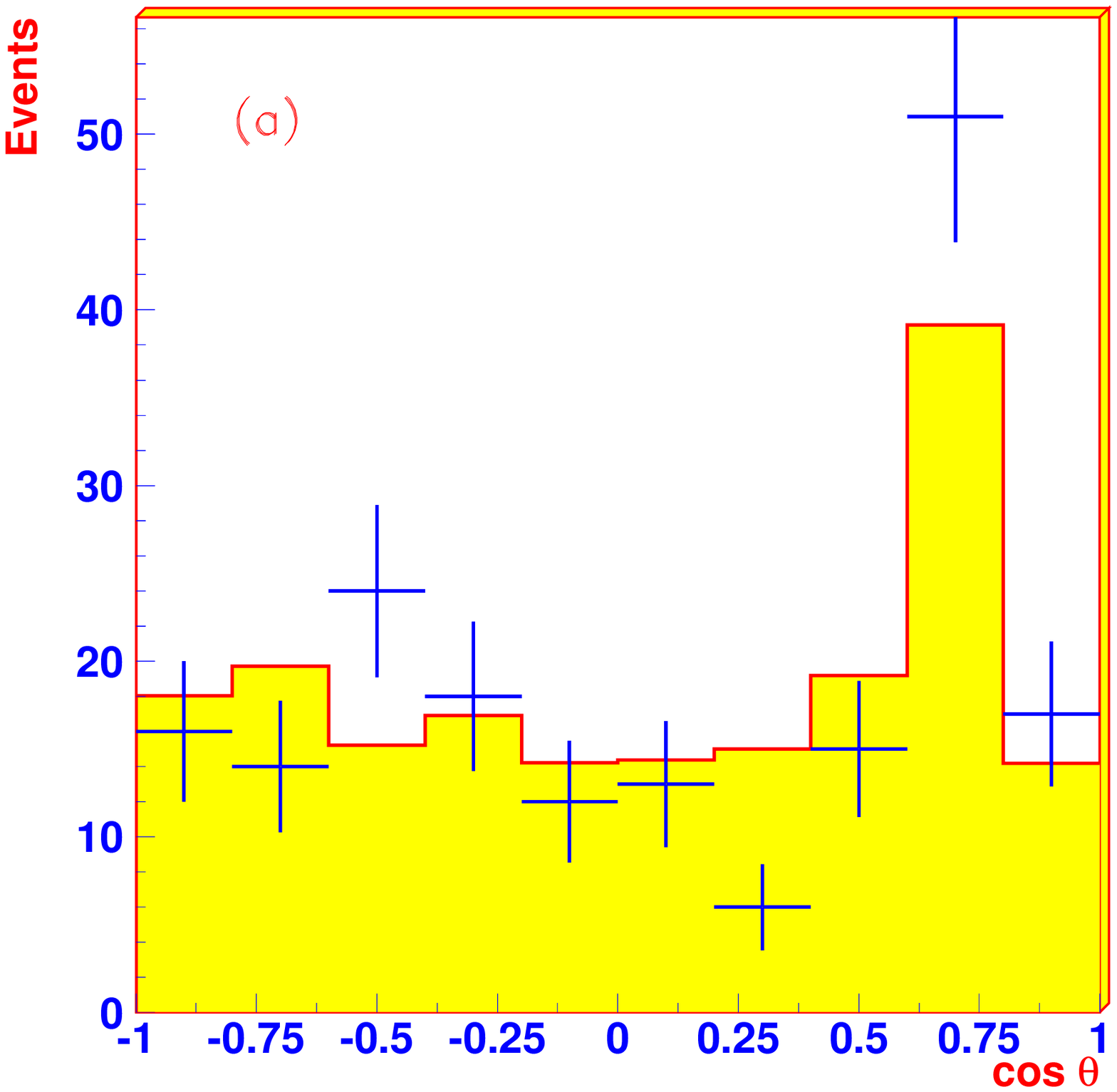,height=1.7in,width=1.7in}}}
\end{flushleft}
\vspace {-1.9in}
\begin{flushright}
{\mbox{\epsfig{file=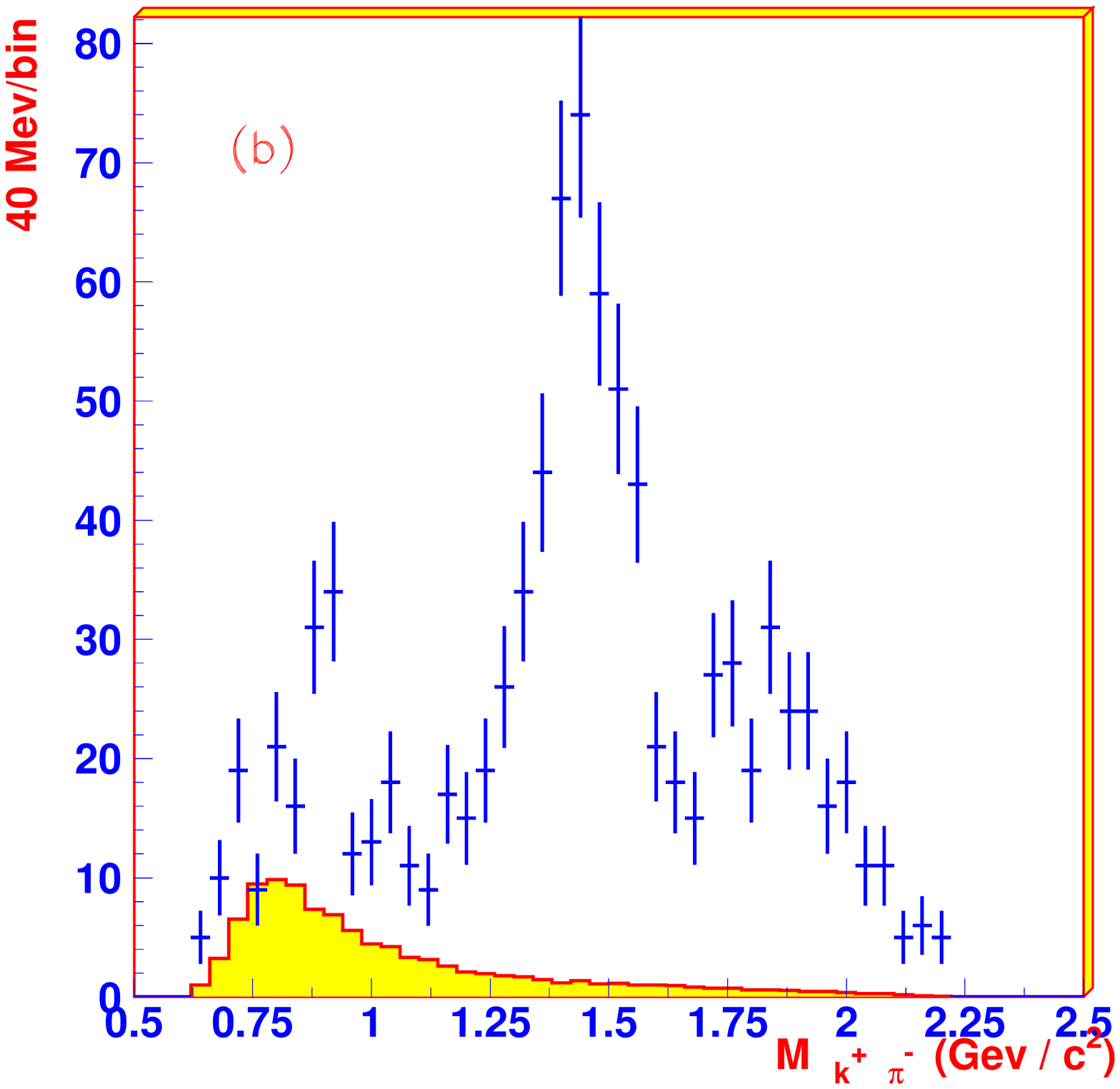,height=1.7in,width=1.7in}}}
\end{flushright}
\begin{flushleft}
{\mbox{\epsfig{file=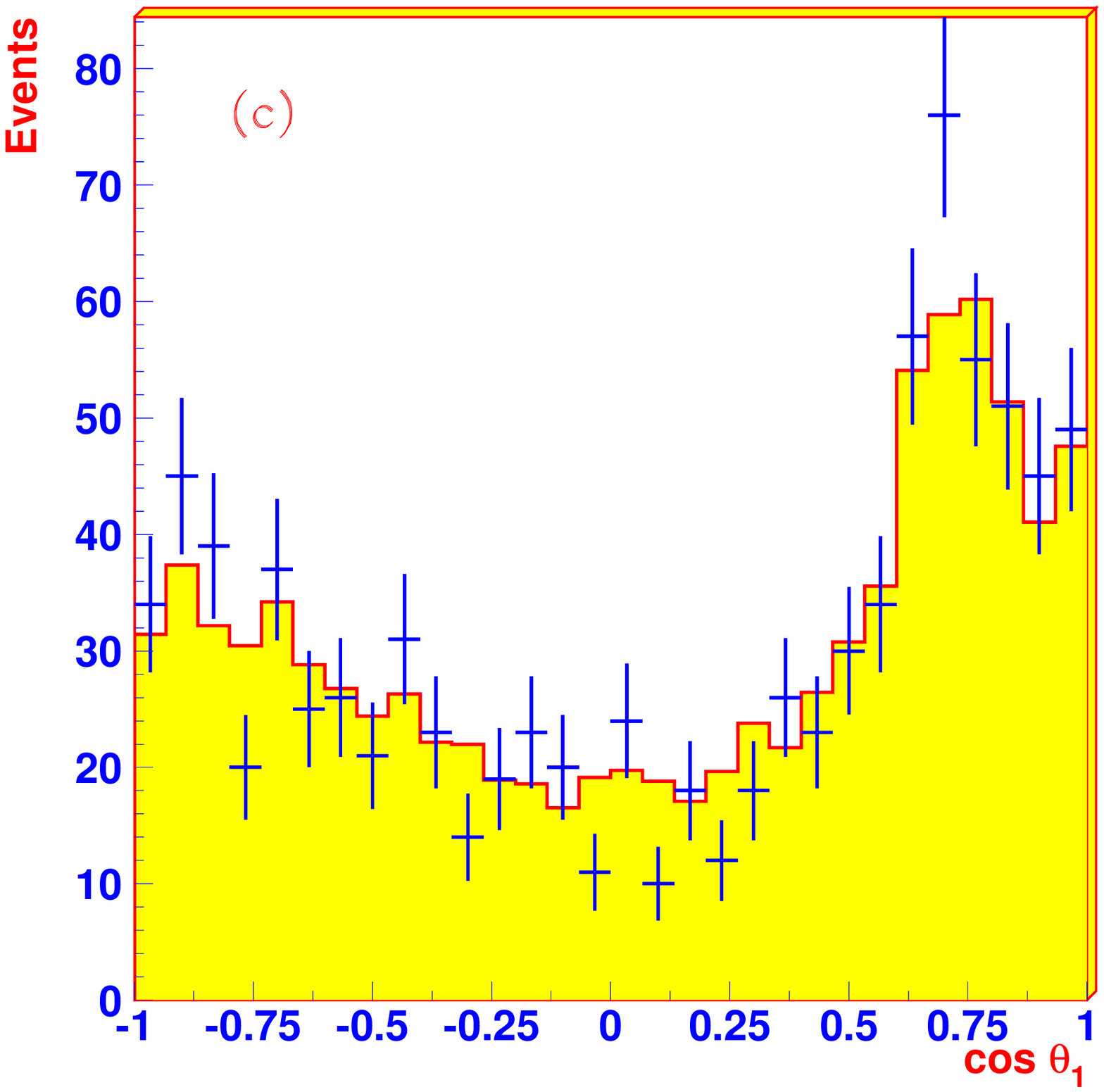,height=1.7in,width=1.7in}}}
\end{flushleft}
\vspace {-1.9in}
\begin{flushright}
{\mbox{\epsfig{file=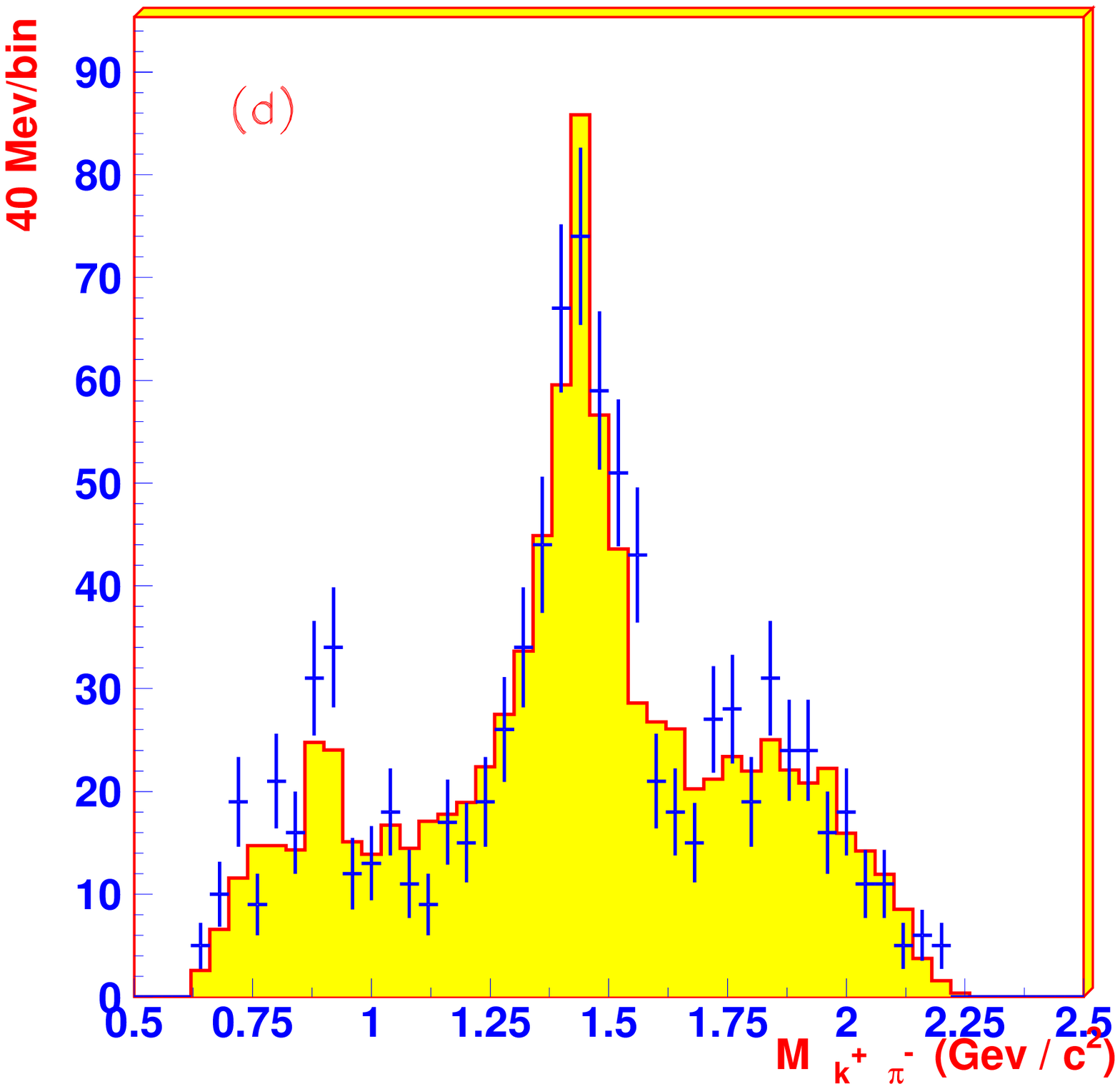,height=1.7in,width=1.7in}}}
\end{flushright}
\caption[]{ (a) The final fit to the angular 
distribution of $K^+$ in $K^+ \pi^-$ rest frame in 
$\kappa$ mass region. The crosses stand for the data and histogram 
 is our final fit. ($M_{K^+ \pi^-} < 1.1 $ GeV ) 
(b) The $\kappa$-particle contribution. The crosses are
the data and the shaded histogram
is the contribution from $\kappa$-particle.
(c) The final global fit to the angular distribution.
$\theta_1$ is the polar angle of $K^+$ particle in the
$\kappa$ rest reference. (d)The final global fit to the invariant
mass spectrum.}
\label{w16}
\end{figure}

In the final fit, $\kappa$, $K^*_0(1430)$, 
$K^*_2(1430)$, $K^*(1410)$,  and the backgrounds from 
$K^*(892)^0 \bar{\kappa}$ and $\bar{K}^*(892)^0 K^0_s$
are considered in the $K^+ \pi^-$ invariant mass
spectrum, $K_1(1270)$ and $K_1(1410)$ are
considered in the $\bar{K}^*(892)^0 \pi^-$ 
invariant mass spectrum. 
In this paper, we report the results using 
the following Breit-Wigner formula for $\kappa$ particle:
$$
BW_{\kappa} = \frac{1}
{m_{\kappa}^2 - s - i m_{\kappa} \Gamma_{\kappa}}
$$
where $\Gamma_{\kappa}$ is a constant.
We first use a $0^{+}$ 
to fit the low mass enhancement, then
we test the statistical significance of its existence.
If we omit it from the fit, the 
log likelihood value  worsened by about 9, which
corresponds to a 3 $\sigma$ statistical significance.
Hence, a resonance with about 3 $\sigma$ statistical 
significance is needed. When the spin-parity 
is changed to $1^-$ or $2^+$, the
log likelihood value becomes slightly worse.
Therefore, its spin-parity cannot be determined 
definitely  using BESI data, due to limited statistics. 
The final fit to the angular distribution of
$\kappa$ in the mass region of $M_{K^+\pi^-}<1.1 $ GeV  
is shown in Fig.\ref{w16}(a) and the $\kappa$ contribution
in $K^+ \pi^-$ mass spectrum is shown in Fig.\ref{w16}(b).
The fits to the angular distribution of the whole mass region
and the $K^+ \pi^-$  invariant mass spectrum
are shown in Fig.\ref{w16}(c) and Fig.\ref{w16}(d)
respectively. \\

The mass and width of the low mass enhancement are
determined to be:
$$
M_{\kappa}=771^{+164}_{-221} \pm 55 MeV/c^2,
\Gamma_{\kappa} = 220^{+225}_{-169} \pm 97 MeV/c^2.
$$
through the scan of mass and width, where the first 
error is statistical and the second is systematic.
The systematic errors
of the mass and width  come from the different
parametrization of the Breit-Wigner formula and different
fit  scheme.
The branching ratio can also be determined to be:
$$
Br(J/\psi \to \bar{K}^*(892)^0 \kappa \to \bar{K}^*(892)^0 K^+ \pi^-)
$$
$$
=(1.26 \pm 0.75 \pm 0.95 ) \times 10^{-4}
$$
where the systematic error is  a combination of the 
contributions from the different parametrization
of the Breit-Wigner formula, different fit scheme, difference
between Monte Carlo and data and the uncertainties of the
$J/\psi$ total number.\\

In summary, by studying the decay of 
$J/\psi \to \bar{K}^*(892)^0 K^+ \pi^- $, 
a low mass enhancement in the $K^+ \pi^-$ 
invariant mass spectrum is observed, which is
not from the phase space effect and backgrounds.
In the partial wave analysis, if we treat the low 
mass enhancement as a resonance, we find that a resonance 
with 3 $\sigma$ statistical significance is needed 
in the fit with the mass and width of 
$771^{+164}_{-221} \pm 55$ MeV/c$^2$ and
$220^{+225}_{-169} \pm 97$ MeV/c$^2$ respectively. 
The branching ratio is   determined as
$(1.26 \pm 0.75 \pm 0.95 ) \times 10^{-4}$.
The mass and width of $\kappa$ are model-dependent. 
Different parametrizations other than simple 
Breit-Wigner formula and 
other forms of the effective vertices 
may give  broader width for $\kappa$. 
\\

We would like to thank Prof. David Bugg, Prof. S. Ishida,
Dr. M. Ishida, Dr. T. Komada, Prof. K. Takamatsu, 
Prof. T.Tsuru, and Prof. K. Ukai  for useful discussions 
on $\sigma$- and $\kappa$-particles.
The BES collaboration thanks the staff of BEPC 
and  computing center for their hard efforts.
This work is supported in part by the National Natural Science Foundation
of China under contracts Nos. 19991480, 10225524, 10225525, 
the Chinese Academy 
of Sciences under contract No. KJ 95T-03, the 100 Talents Program of CAS 
under Contract Nos. U-24, U-25, and the Knowledge Innovation Project of 
CAS under Contract Nos. U-602, U-34(IHEP); by the National Natural Science 
Foundation of China under Contract No.10175060(USTC).
\\

%\vspace{-0.5cm}

\end{document}